\documentclass[
reprint,
superscriptaddress,
amsmath,amssymb,
aps,
pra,
floatfix,
]{revtex4-2}

\usepackage{booktabs}
\usepackage{amsmath,amsthm,amsfonts,amssymb,extarrows}
\usepackage{lipsum}
\usepackage{arydshln}
\usepackage{multirow}
\usepackage{graphicx}
\usepackage[detect-all]{siunitx} 
\usepackage{xcolor}
\definecolor{mBlue}{RGB}{51, 77, 167}

\usepackage[colorlinks=true,citecolor=blue,linkcolor=red,urlcolor=blue]{hyperref}
\usepackage{xcolor}

\usepackage{romannum}
\usepackage{romannum}
\newcommand{\ion}[2]{$\mathrm{#1\,\uppercase\expandafter{\romannumeral #2}}$}

\begin{document}

    \title{\Large Laboratory Measurements of \ion{Ca}{19} Dielectronic Recombination Satellites}
	

	\author{Filipe Grilo}\email{f.grilo@campus.fct.unl.pt} 
	\affiliation{Laboratory of Instrumentation, Biomedical Engineering and Radiation Physics (LIBPhys-UNL), Department of Physics, NOVA School of Science and Technology, NOVA University Lisbon, 2829-516 Caparica, Portugal}

    \author{Marc Botz}
	\affiliation{Max-Planck-Institut f\"ur Kernphysik, Saupfercheckweg 1, 69117 Heidelberg, Germany}
	
	\author{Chintan Shah}%
	\affiliation{NASA Goddard Space Flight Center, 8800 Greenbelt Rd, Greenbelt, MD 20771, USA}
	\affiliation{Max-Planck-Institut f\"ur Kernphysik, Saupfercheckweg 1, 69117 Heidelberg, Germany}
    \affiliation{Department of Physics and Astronomy, Johns Hopkins University, Baltimore, MD 21218, USA}
	
    \author{Thomas~Pfeifer}
    \affiliation{Max-Planck-Institut f\"ur Kernphysik, Saupfercheckweg 1, 69117 Heidelberg, Germany}	

	\author{Jos\'e R. {Crespo L\'opez-Urrutia}}
	\affiliation{Max-Planck-Institut f\"ur Kernphysik, Saupfercheckweg 1, 69117 Heidelberg, Germany}
	
	\author{Pedro Amaro}\email{pdamaro@fct.unl.pt}
	\affiliation{Laboratory of Instrumentation, Biomedical Engineering and Radiation Physics (LIBPhys-UNL), Department of Physics, NOVA School of Science and Technology, NOVA University Lisbon, 2829-516 Caparica, Portugal}


\begin{abstract}

We report measurements of the K$\alpha$ emission from the astrophysically very abundant \ion{Ca}{19} (He-like ion) and its satellite lines resonantly excited by dielectronic recombination (DR). We achieve an electron-energy resolution of 8~eV in a cryogenic electron beam ion trap, and determine the energies of the exciting electrons and the emitted photons up to the KLn ($n\le 8$) manifold with 0.05\% and 0.1\% respective uncertainties. For the KLL satellites, energies agree very well with our predictions using the Flexible Atomic Code (FAC) and previous state-of-the-art calculations. Our calculations also agree with our experimental direct excitation cross-sections for K$\alpha$ within their 10\% uncertainty. We extract DR coefficient rates and find good agreement with values tabulated in the OPEN-ADAS database. As an application, we experimentally benchmark \ion{Ca}{19} atomic data used to model high-temperature astrophysical plasmas by comparing FAC synthetic spectra with recent XRISM observations revealing the contributions of DR satellites to the \ion{Ca}{19} lines.

\end{abstract}

\date{\today} 

\maketitle

\section{Introduction}
\label{intro}

The advent of high-resolution X-ray spectrometers onboard space missions brought an urgent need for accurate laboratory data. While early X-ray space missions, like UHURU \citep{Giacconi1971}, Ariel V \citep{Smith1976} and Heo-1 \citep{Wood1984}, provided only broadband spectra, later ones equipped with higher-resolution grating spectrometers such as Chandra \citep{Canizares2000,Brinkman2000} and XMM-Newton \citep{Herder2001} enabled more detailed plasma diagnostics \citep{Paerels2003}. Results (see reviews by ~\cite{Paerels2003, Ezoe2021}) included, e.~g., electron temperature \citep{Peterson2003} and density distributions \citep{Ness2001}, bulk Doppler velocities of outflows \citep{Kaspi2002} and estimates of turbulence velocities \citep{Kaspi2002}. The analysis of such observations with spectral models crucially depends on the accuracy of the underlying atomic data.
Recently, the Soft X-ray Spectrometer (SXS) microcalorimeter onboard the ill-fated Hitomi satellite acquired high-resolution spectra of the Perseus cluster that allowed to characterize the turbulent motion at its center \citep{Takahashi2016} based on the broadening and Doppler shift of transitions of He-like ions. However, differences in theoretical atomic data used in the spectral codes AtomDB/APEC \citep{Foster2012}, SPEX \citep{Kaastra1996}, and CHIANTI \citep{Young2016} resulted in large discrepancies of up to 17\% \citep{Aharonian2018} for the iron abundance derived from the He-like $z/w$ intensity ratio. Therefore, benchmark experiments are required to solve this problem. Nonetheless, the excellent resolution of Hitomi showed the overwhelming advantages of microcalorimeter-based X-ray astrophysics. A follow-up mission, the X-Ray Imaging and Spectroscopy Mission (XRISM) equipped with the Resolve microcalorimeter, is now operational, and has already delivered groundbreaking results, e.~g., on the velocity distribution and thermal properties of the supernova remnant N123D \citep{10.1093/pasj/psae080}, the bulk-gas motion in the Centaurus galaxy cluster \citep{ Audard2025}, the accretion flow of the active galactic nucleus NGC 4151 \citep{Audard2024}, stellar winds in the binary system Cygnus X-3 \citep{Audard2024b} and the ionization state of the supernova remnant Sagittarius A East \citep{Audard2025b}. 

Unfortunately, the gate valve in front of the XRISM detector malfunctioned in orbit, and observations have to take place through a beryllium window installed on it for pre-launch calibration. It strongly absorbs soft X-rays below 2\,keV, hindering crucial diagnostics of e.~g., the Fe L-shell complex \citep{Audard2025}. Thus, spectral diagnostics work only at higher energies, and rely on H-like and He-like K$\alpha$ lines of S, Ca, and Fe and their satellites \citep{10.1093/pasj/psae080}. One example is the Fe He-like K$\alpha$ transition used to analyze the bulk motion of the plasma at the Centaurus galaxy cluster \citep{Audard2025}. The present experimental study benchmarks the K$\alpha$ satellites of He-like  \ion{Ca}{19}, which belong to the strongest spectral features seen by XRISM. 
 
Closed-shell He-like ions display very strong $n=2\rightarrow n=1$ transitions, labeled following the \cite{Gabriel1969} notation. The most studied ones are the resonance line $w$ ($1s 2p~^1\mbox{P}_1- 1s^2~^1\mbox{S}_0$), the intercombination line--composed of two transitions, $x$ ($1s 2p~^3\mbox{P}_2 - 1s^2~^1\mbox{S}_0$) and $y$ ($1s 2p~^3\mbox{P}_1 - 1s^2~^1\mbox{S}_0$)--as well as the forbidden line $z$ ($1s 2s ^3\mbox{S}_1 - 1s^2 ~ ^1\mbox{S}_0$). They are often accompanied by less intense Li-like satellite lines excited through dielectronic recombination (DR). Since the pioneering work of \cite{massey1942} and \cite{burgess1964} showed the importance of DR for plasma emissivity, this atomic process was intensively studied both theoretically \citep{Kato1997, Sardar2018, Yerokhin2018, Rui2023} and experimentally \citep{Knapp1989, Hahn1989a, Andersen1989, Beiersdorfer1990,Beiersdorfer1992,Ali1991, Beiersdorfer1996, Shlyaptseva1998,Bitter2003,Brown2006,Watanabe2007,Beilmann2009,Kavanagh2010,Yao2010,Brandau2012,Beilmann2011,Beilmann2013,Shah2015,Shah2016,Shah2021,Amaro2017,Orban2024,Shah_2025}. A compilation of $1s2\ell 2\ell'$ satellite lines is given by \cite{Azarov2023}. However, laboratory data on \ion{Ca}{19} is scarce, stemming from Tokamak and accelerator experiments \citep{Rice2014, Rice2015, Suleiman1994}, and there is no comprehensive study about DR KL$n$ energies and cross-sections.

In this work, we report measurements of the DR KL$n$ satellite emissions and the related collisional and resonant excitation of He-like \ion{Ca}{19} taken with an electron beam ion trap (EBIT). 
The high resolution of the electron-beam energy of FLASH-EBIT allowed us to accurately determine DR KL$n$ electron and photon DR energies up to the $n=8$ manifold through a two-dimensional fit. Relative collisional cross sections for DR and non-resonant excitation were also measured. These results, and the rate coefficients calculated from them, are in very good agreement with state-of-the-art calculations obtained with the Flexible Atomic Code (FAC) calculations and the values tabulated in the OPEN-ADAS database. Moreover, we analyze the contribution of DR satellites to the Ca K$\alpha$ emissions using FAC collisional-radiative model predictions and compare them with recent XRISM spectra \citep{Audard2025}.

\section{Experimental setup}
\label{Experi_setup}
\begin{figure*}[th!]
 \centering
\includegraphics[clip=true,width=1.0\textwidth]{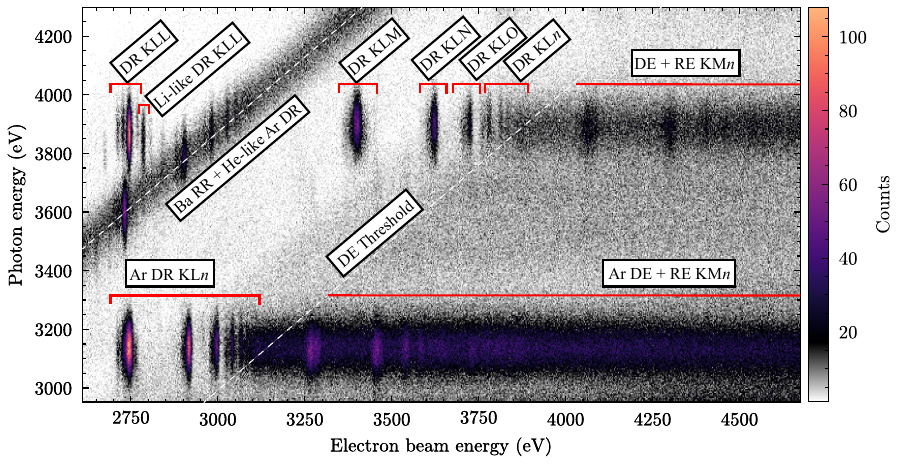}
\caption{Photon yield as a function of electron-beam energy and photon energy for a slow scan across the DR KL$n$ structure of He-like Ca, as well as its DE and RE. A diagonal band of emission results from DR in He-like Ar, as well as L-shell RR into He-like Ar and N-shell RR into Ni-like Ba ions.}
\label{figs:1}
\end{figure*}
FLASH-EBIT \citep{Epp2007,Epp2010} was used to produce and probe He-like ions of Ca at the Max-Planck-Institut f\"ur Kernphysik (MPIK) in Heidelberg. A solid sample of the compound Ca(TMHD)$_2$ (calcium bis(2,2,6,6-tetramethyl-3,5-heptanedionate)) was heated up to 90~\textdegree C in the differentially pumped injection system of the device. The evaporated molecular beam enters the central trap region of the EBIT through narrow apertures, where the injected molecules are dissociated by an electron beam compressed by a 6~T magnetic field to a radius of tens of microns. The beam efficiently ionizes the resulting Ca atoms, and traps radially the generated HCI. 

The electron-beam energy follows from a sawtooth acceleration potential similar to \citep{Grilo2024}, between 2500 eV and 4700 eV, instead of a breeding period at a set energy to produce specific ion species \citep{Shah_2019}. We chose this scheme to maximize the time used for probing the ions, increasing the acquired statistics. This point was vital, since the injection of the Ca compound was not very efficient. Photon emission following the electron-ion interactions was recorded using a silicon-drift detector (SDD) with a photon-energy resolution of around 140 eV of FWHM at around 3800 eV mounted at 90\textdegree~ relative to the electron-beam axis. The energy of each photon was measured as a function of the electron-beam energy, yielding two-dimensional spectral plots (for more details, see \citep{Grilo2024}).

In the scanned electron-energy range, we could observe the entire DR KL$n$ series of He-like Ca converging toward the K$\alpha$ emission. This emission mainly arises from direct excitation (DE), with a small contribution of resonant excitation (RE), along with some Li-like DR KLL resonances. The ionization threshold of Li-like Ca is approximately 1160 eV, much lower than that of the He-like ion around 5130 eV. This means that within the scan range a charge-state distribution with around 80\% of He-like and 20\% Li-like is established. A contamination by Ar ions of nearly three times the concentration as the Ca ions was present, likely due to an air (1\% Ar concentration) leak in the injection system, as the Ar lines were only visible when the injection was in operation. Nonetheless, the Ar and Ca K$\alpha$ emissions are 600 eV apart, and their interference is mostly negligible. The background between Ar and Ca K$\alpha$ consists of Ar K$\beta$ ($\approx3.7$~keV) and M-shell transitions of Ni-like and nearby charge-states of W ($\approx3.5-3.6$~keV) \citep{Clementson2010}, with Ar partly blending with Ca K$\alpha$ (see Sec.~\ref{sec:cross}). Moreover, a radiative recombination (RR) band closely coinciding with that of He-like Ar into the L-shell is clearly visible, crossing through the Ca K$\alpha$ region between its KLL and the KLM features. However, as discussed in more detail in Appendix~\ref{sec:RRveri}, this feature is predominantly due to N-shell RR emission into ions of Ba, with only a small contribution of RR into He-like Ar. However, it remains spectrally distinct from the Ca DR resonances and does not overlap with them.

Scans were performed with very different ramping speeds, with periods ranging from 500~s for slow and 30~ms for fast scans, respectively. During slow scans the charge-state distribution evolves and the population of He-like Ca changes, but better electron beam energy resolution is reached. In the fast scans, this dynamic behavior is suppressed and a stable average He-like population is established, but energy resolution is lower. Therefore, we use slow scans for the determination of resonant energies, while excitation cross-sections were obtained from fast scans. 

The SDD spectrum was calibrated by fitting well-known soft X-ray lines for a wide range of energies. The electron beam energy calibration was made by matching two isolated DR resonances, KLL and KLO, with their respective FAC values, indicated in Tables \ref{tab:CaLiComp} and \ref{tab:Cafit}). More details on the calibration procedure can be found in Appendix \ref{sec:calibration}.

%
\section{Calculations}
\label{sec:calc}
The Flexible Atomic Code (FAC) \citep{gu2008} was used to predict electronic levels and transition rates (radiative and autoionizing) of the ions of interest. The energy levels, autoionization $A^{{a}}$ as well as radiative rates $A^{{r}}$ are calculated in FAC within the configuration interaction (CI) approach. Due to the unidirectionality of the electron beam, photon emission is polarized and anisotropic. For a direct comparison with the experimental data, we correct the theoretical cross section for the transition multipolarity and polarization angular distribution. These corrections were calculated with the polarization module of FAC, and can be expressed for an observation angle of 90\textdegree ~as
\begin{equation}
    \Gamma (\theta = 90 ^{\circ}) = \frac{3}{3 \mp P},
\end{equation}
where $P$ is the degree of polarization, while the plus and minus signs are for M1 and E1 transitions, respectively.
We evaluated resonant recombination processes, such as DR and RE, according to the isolated-resonance approximation, whereby these processes are considered without quantum interference effects between the non-resonant  RR and the DR resonances themselves leading from the same initial to the same final state. Both DR and RE processes are described as a two-step process of an inverse autoionization process followed by photon emission (DR case) or an autoionizing decay. In this case, the resonance strength (integral of the cross section) of each resonance is given by
\begin{eqnarray}
 S_{idf}^{x} &=& \int_{0}^{\infty} \sigma^{x}_{idf}(E_e) d E_e \nonumber \\
 			 &=& \frac{\pi^2 \hbar^3}{m_e E_{id}} \frac{g_d}{2 g_i} \frac{A^{a}_{di} A^{y}_{df}}{\sum_{i'} A^{a}_{di'} + \sum_{f'} A^{r}_{df'}},
\end{eqnarray}
where $ \sigma^{x}_{idf}(E_e)$ is the cross section of DR ($x=$DR, $y=$r), or RE ($x=$RE, $y=$a) as a function of the free-electron kinetic energy $E_{e}$ for the process of inverse-autoionization of the state $i$ towards an intermediate state $d$ followed by a photon decay (DR) or a autoionization decay (RE) to the final state $f$. $E_{id}$ is the resonant energy of the electron$-$ion recombination between initial $i$ and intermediate $d$ states, with respective statistical weights $g_{{i}}$ and $g_{{d}}$. $m_{{e}}$ is the electron mass. 

For the DR calculations we include the configurations of the ground state as well as singly and doubly excited states $1s n \ell$ and $2 \ell n \ell '$ up to $n\le15$. 
The DE and RR cross sections are obtained using the distorted-wave approximation from FAC including singly excited configurations up to $n_{\mbox{\small max}}=10$. We also calculate higher $n$ excitations to include their cascade contributions to the K$\alpha$ emission.

\begin{figure*}[th!]
 \centering
\includegraphics[clip=true,width=1.0\textwidth]{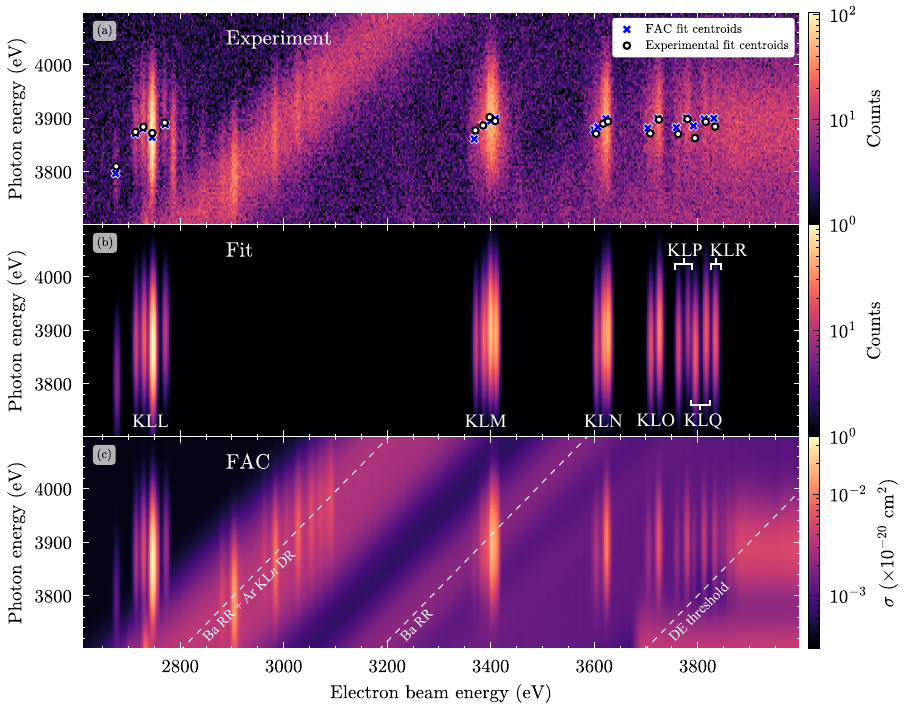}
\caption{Fit procedure for the DR emission plots. (a) Experimental results. Blue crosses mark the fit centroids of the FAC data; black crosses those of the experimental data. (b) Plot of the experimental fit. (c) Plot of DR KL$n$ FAC calculations convoluted with the experimental widths. The DR structures of He-like Ar as well as the RR and DE features of both Ca, Ar and Ba were also included for comparison with the experiment.}
\label{figs:fitCa}
\end{figure*}

In order to identify features not belonging to the He-like ion, we 
calculate the Li-like structure with a set of configurations from the ground state $1s^2 2s$, singly excited and doubly excited state configurations $1s^2 n \ell$ and $1s 2 \ell n \ell '$. Since the Li-like DR emissions are weak, $n_{\mbox{\small max}} = 5$ was used. 
A complete map of the expected theoretical cross-sections from DR, DE and RR is shown in subplot (c) of Figure \ref{figs:fitCa}.
%
\section{Result analysis and Discussion}
\label{sec:disc}

\begin{table*}[th!]
\centering
\caption{Fitted experimental photon energies (in eV) of KLL satellite transitions of He-like Ca labeled according to \cite{Gabriel1972}’s notation. Results are compared with our FAC calculations and predictions obtained with the MZ and AUTOLSJ codes presented in \cite{Kato1997} and the CI+QED method by \cite{Yerokhin2018}. E$\gamma$-FAC values show photon energies fitted to the FAC data convoluted with the experimental width; FAC values denote the calculated transition energies. Since most lines are blended, we compare blended values from experiment and theory, with the relative difference shown in parentheses.}
\begin{tabular}{cccccccccc}
\toprule
Label & Transition & E$_e$-exp			&	E$_e$-FAC	&	E$\gamma$-exp			& E$\gamma$-FAC &	FAC &  AUTOLSJ & MZ & CI+QED \\ \midrule
\multirow{2}{*}{$p$} & \multirow{2}{*}{$1s 2s^2 ~^2\mbox{S}_{1/2} \rightarrow 1s^2 2p ~^2\mbox{P}_{3/2}$}  
    & \multirow{4}{*}{2675.3$^\text{a}$ } 
    & \multirow{4}{*}{2675.30} 
    & \multirow{4}{*}{\shortstack{3806 \\ $\pm 6$}} 
    & \multirow{4}{*}{3791.8 } 
    & 3796.9 & 3800.7 & 3797.3 & 3797.6 \\
 &  &  &  & &  & ($-0.08\%$) &  & ($-0.07\%$) & ($-0.06\%$) \\
\multirow{2}{*}{$o$} & \multirow{2}{*}{$1s 2s^2 ~^2\mbox{S}_{1/2} \rightarrow 1s^2 2p ~^2\mbox{P}_{1/2} $ }
    &  &  & &  & 3791.8 & 3794.7 & 3792.1 & 3792.5 \\
 &  &  &  & &  & ($-0.22\%$) & ($-0.14\%$) & ($-0.21\%$) & ($-0.20\%$) \\ \midrule
\multirow{2}{*}{$r$}	&	 \multirow{2}{*}{$1s 2s 2p ~^2\mbox{P}_{1/2}	\rightarrow	 1s^2 2s ~^2\mbox{S}_{1/2} $}	
    & \multirow{2}{*}{\shortstack{2712.4  \\ $\pm 1.3$}} 
    & \multirow{2}{*}{2713.74} 
    & \multirow{2}{*}{\shortstack{3870 \\ $\pm 3$}} 
    & \multirow{2}{*}{3871.1 } 
    & 3870.8 & 3874.4 & 3871.2 & 3871.2 \\
	&		 &		&				&		&				&		&	 ($0.04\%$)	&	 	&	 	\\ \midrule
\multirow{2}{*}{$t$} &  \multirow{2}{*}{$1s 2s 2p ~^2\mbox{P}_{1/2} \rightarrow 1s^2 2s ~^2\mbox{S}_{1/2}$ }
    & \multirow{4}{*}{\shortstack{2727.6  \\ $\pm 1.3$}} 
    & \multirow{4}{*}{2727.93} 
    & \multirow{4}{*}{\shortstack{3880  \\ $\pm 2$}} 
    & \multirow{4}{*}{3880.5 } 
    & 3884.9 & 3888.0 & 3884.8 & 3884.9 \\
    &  & & & & & ($0.07\%$) & ($0.15\%$) & ($0.07\%$) & ($0.07\%$) \\
\multirow{2}{*}{$s$} &  \multirow{2}{*}{$1s 2s 2p ~^2\mbox{P}_{3/2} \rightarrow 1s^2 2s ~^2\mbox{S}_{1/2}$ }
    & & & & & 3885.9 & 3889.5 & 3885.9 & 3885.9 \\
    &  & & & & & ($0.10\%$) & ($0.19\%$) & ($0.10\%$) & ($0.10\%$) \\ \midrule

\multirow{2}{*}{$j$} &  \multirow{2}{*}{$1s 2p^2 ~^2\mbox{D}_{5/2} \rightarrow 1s^2 2p ~^2\mbox{P}_{3/2}$ }
    & \multirow{4}{*}{\shortstack{2744.8  \\ $\pm 1.3$}} 
    & \multirow{4}{*}{2745.98} 
    & \multirow{4}{*}{\shortstack{3868  \\ $\pm 1$}} 
    & \multirow{4}{*}{3864.3 } 
    & 3862.3 & 3865.7 & 3862.2 & 3862.3 \\
    &  & & & & & ($-0.12\%$) & ($-0.03\%$) & ($-0.12\%$) & ($-0.12\%$) \\
\multirow{2}{*}{$k$} & \multirow{2}{*}{ $1s 2p^2 ~^2\mbox{D}_{3/2} \rightarrow 1s^2 2p ~^2\mbox{P}_{1/2}$ }
    & & & & & 3866.9 & 3870.4 & 3866.8 & 3866.8 \\
    &  & & & & & ($-0.003\%$) & ($0.04\%$) & ($-0.005\%$) & ($-0.005\%$) \\ \midrule

\multirow{2}{*}{$n$} &  \multirow{2}{*}{$1s 2p^2 ~^2\mbox{S}_{1/2} \rightarrow 1s^2 2p ~^2\mbox{P}_{1/2}$ }
    & \multirow{4}{*}{\shortstack{2768.9  \\ $\pm 1.3$}} 
    & \multirow{4}{*}{2771.21} 
    & \multirow{4}{*}{\shortstack{3887  \\ $\pm 4$}} 
    & \multirow{4}{*}{3888.3 } 
    & 3893.0 & 3897.0 & 3892.1 & 3892.1 \\
    &  & & & & & ($0.05\%$) & ($0.15\%$) & ($0.03\%$) & ($0.03\%$) \\
\multirow{2}{*}{$m$} &  \multirow{2}{*}{$1s 2p^2 ~^2\mbox{S}_{1/2} \rightarrow 1s^2 2p ~^2\mbox{P}_{3/2}$ }
    & & & & & 3887.9 & 3890.9 & 3887.0 & 3887.0 \\
    &  & & & & &  &  &  &  \\
\bottomrule
\end{tabular}
\label{tab:CaLiComp}
\begin{flushleft}
\footnotesize
\tablenotemark{}{Used for the electron beam energy axis calibration. }
\end{flushleft}
\end{table*}

\subsection{Experimental energy positions}

\begin{table*}[th!]
\centering
\caption{Same as Table~\ref{tab:CaLiComp} but for the KL$n$ ($3 \le  n \le 8$) satellite transitions. Since all resolved peaks are a blend of several resonances, we identify them by the leading one.   
}
\begin{tabular*}{\linewidth}{ @{\extracolsep{\fill}} cccccc}
\toprule
	&	Leading transition	&	E$_e$-exp			&	E$_e$-FAC	&	E$\gamma$-exp			&	E$\gamma$-FAC 	\\	\midrule
KLM	&	$1s2s3s ~^2\mbox{S}_{1/2}	 \rightarrow 	 1s^2 3p ~^2\mbox{P}_{3/2}$	&	3368.6	$ \pm $	1.0	&	3368.20	&	3873	$ \pm $	3	&	3861.4	\\
	&	$1s2p3s ~^2\mbox{P}_{1/2}	 \rightarrow 	 1s^2 3s ~^2\mbox{S}_{1/2}$	&	3383.5	$ \pm $	1.0	&	3385.88	&	3882	$ \pm $	2	&	3887.1	\\
	&	$1s2p3p ~^2\mbox{D}_{5/2}	 \rightarrow 	 1s^2 3p ~^2\mbox{P}_{3/2}$	&	3396.0	$ \pm $	1.0	&	3400.34	&	3898	$ \pm $	2	&	3895.0	\\
	&	$1s2p3d ~^2\mbox{F}_{7/2}	 \rightarrow 	 1s^2 3d ~^2\mbox{D}_{5/2}$	&	3406.0	$ \pm $	1.0	&	3409.16	&	3890	$ \pm $	2	&	3899.4	\\
KLN	&	$1s2s4p ~^2\mbox{P}_{1/2}	 \rightarrow 	 1s^2 4s ~^2\mbox{S}_{1/2}$	&	3601.4	$ \pm $	1.2	&	3600.02	&	3867	$ \pm $	3	&	3878.3	\\
	&	$1s2p4p ~^2\mbox{D}_{5/2}	 \rightarrow 	 1s^2 4p ~^2\mbox{P}_{3/2}$	&	3615.2	$ \pm $	1.2	&	3606.99	&	3886	$ \pm $	2	&	3883.9	\\
	&	$1s2p4d ~^2\mbox{F}_{7/2}	 \rightarrow 	 1s^2 4d ~^2\mbox{D}_{5/2}$	&	3624.5	$ \pm $	1.2	&	3622.47	&	3890	$ \pm $	2	&	3899.1	\\
KLO	&	$1s2s5p ~^2\mbox{D}_{3/2}	 \rightarrow 	1s^2 5p ~^2\mbox{P}_{3/2}$	&	3706.2	$ \pm $	1.3	&	3703.84	&	3867	$ \pm $	3	&	3881.7	\\
	&	$1s2p5d ~^2\mbox{F}_{7/2}	 \rightarrow 	1s^2 5d ~^2\mbox{D}_{5/2}$	&	3723.0$^\text{a}$	&	3722.97	&	3893	$ \pm $	2	&	3899.8	\\
KLP	&	$1s2p6p ~^4\mbox{D}_{5/2}	 \rightarrow 	1s^2 6p ~^2\mbox{P}_{3/2}$	&	3760.0	$ \pm $	1.4	&	3870.66	&	3866	$ \pm $	3	&	3882.6	\\
	&	$1s2p6d ~^2\mbox{F}_{7/2}	 \rightarrow 	1s^2 6d ~^2\mbox{D}_{5/2}$	&	3778.1	$ \pm $	1.4	&	3898.38	&	3894	$ \pm $	3	&	3900.5	\\
KLQ	&	$1s2p7p ~^2\mbox{D}_{3/2}	 \rightarrow 	1s^2 7p ~^2\mbox{P}_{3/2}$	&	3792.4	$ \pm $	1.5	&	3863.01	&	3859	$ \pm $	4	&	3885.8	\\
	&	$1s2p7d ~^2\mbox{F}_{7/2}	 \rightarrow 	1s^2 7d ~^2\mbox{D}_{5/2}$	&	3813.2	$ \pm $	1.5	&	3812.80	&	3889	$ \pm $	3	&	3899.9	\\
KLR	&	$1s2p8d ~^2\mbox{F}_{7/2}	 \rightarrow 	1s^2 8d ~^2\mbox{D}_{5/2}$	&	3831.3	$ \pm $	1.5	&	3831.44	&	3880	$ \pm $	3	&	3898.8	\\
\bottomrule
\end{tabular*}\\
\vspace{1ex}
\begin{flushleft}
\footnotesize
\tablenotemark{}{Used for the electron beam energy axis calibration. }
\end{flushleft}
\label{tab:Cafit}
\end{table*}

The slow scan yielded an electron beam energy spread with a FWHM of approximately 8 eV at 2700 eV. We extract from it for each of the resolved resonances of the KL$n$ feature the experimental electron and photon energy centroids by means of a fit of linear combinations of two-dimensional Gaussian functions using the Minuit2 library of the ROOT analysis package \citep{Antcheva2009} with standard $\chi^2$-minimization. 

Determining how many Gaussian functions are fitted is difficult, since tens of resonances can be blended in the data. For this work, we employed the $K$-means clustering method \citep{lloyd1982} to group the theoretical resonances into clusters. The number of clusters $n_c$ can be defined by how many features can be included without the minimum separation between clusters being less than the experimental width. The cluster centroids are then used as the initial parameters of the fit of $n_c$ 2D-Gaussians to the theory values (convoluted with the experimental widths of the photon and electron energy). The results of this fit are later used as initial parameters for the experimental fit. This ensures consistency between the fits to the theoretical and experimental results.

Applying this procedure to the KL$n$ DR features allowed us to identify and fit resonances up to KLR ($n=8$). A comparison of the experimental and fit results is plotted in Figure \ref{figs:fitCa}. Subfigure (a) shows the experimental data, superimposed with the centroids obtained with the fit procedure (black circles). Fitted the theoretical values convoluted by the experimental resolutions are compared, with the corresponding centroids also displayed as blue crosses. In general, the independently calibrated experimental data and the theoretical centroids mostly agree within the quoted uncertainties. As expected, there is better agreement at lower electron energies than at higher ones, since a growing number of blended resonances with $n$ challenges the clustering method for peak fitting as well as the accuracy of the predictions. 
 
The diagonal emission feature between 2800~eV and 3200~eV is due to RR of He-like Ar, and displays DR resonances of this ion that were not fitted; KLL resonances of Li-like Ca resonance below 2800 eV were also excluded. Subfigures (b) and (c) show the functions obtained from the fitting procedure for the experimental, and the convoluted FAC data. The calculations show more discernible structures at the higher energies of subfigure (c) than in subfigure (b). The observed intensities relative to the strongest KLL peak seem to be generally more intense than predicted. This could result from the clear changes of the charge-state distribution taking place over the slow-scan range.
The centroids obtained from the experimental and theoretical data fits are compiled in Tables \ref{tab:CaLiComp} and \ref{tab:Cafit} for KLL and KL$n$ ($3 \le n \le 8$), respectively. Listed uncertainties take into account both the uncertainty in photon and electron energy calibration (see Appendix~\ref{sec:calibration}), as well as the statistical uncertainty obtained from the fits.

While the KLL satellites of Ca have been theoretically studied by a few groups, currently no experimental or theoretical studies are reported for KL$n$ ($3 \le n \le 8$). Table~\ref{tab:CaLiComp} compares the observed KLL satellite lines with the notation of \cite{Gabriel1972} and current state-of-the-art calculations, ordered by the electron energy of the resonances. As most of the experimental peaks have two blended satellites, each experimental value is compared with both theoretical values. Five experimental lines were observed, with the corresponding excited configurations being $1s 2s^2$ ($p+o$), $1s 2s 2p$ ($r$, $t+s$), and $1s 2p^2$ ($j+k$, $m+n$).
These values were compared with the transition energies given by    \cite{Kato1997} and \cite{Yerokhin2018}, and our FAC calculations. Values in parentheses in Table~\ref{tab:CaLiComp} indicate the relative difference between the theoretical value and experimental values. Transition energies included in \cite{Kato1997} were calculated with  AUTOLSJ \citep{Cornille1993} package and MZ perturbation method \citep{VAINSHTEIN197849}. The AUTOLSJ package uses a scaled Thomas-Fermi-Dirac potential to estimate the radial wavefunctions with the SUPERSTRUCTURE code \citep{Eissner1974}, where scaling parameters are optimized to the resulting energies. AUTOLSJ needs the level mixings from the previous code to calculate the autoionization rates.
Safronova applies the MZ perturbation method with a semi-relativistic $Z$-expansion of atomic parameters in powers of nuclear charge $Z$ to calculate the energy levels taking into account relativistic effects and electron correlation.
The calculations from \cite{Yerokhin2018} with a fully relativistic CI method include the frequency-dependent Breit interaction, relativistic nuclear recoil, and quantum electrodynamics effects, such as self-energy, vacuum polarization and other higher-order contributions.

All predicted lines fairly well agree with the experimental results, deviating by less than 0.16\% from the predictions of \cite{Yerokhin2018}. Only the two-electron-one-photon (TEOP) transitions $p$ and $o$ deviate from the most recent predictions by 0.20\% to 0.40\%, respectively; these seem to be underestimated in all predictions. Excellent agreement was found for the $r$, $j+k$ and $m+n$ lines ($j+k$ being the strongest KLL resonance), agreeing with most calculations within 0.05\%. FAC results are closest to those of Safronova and Yerokhin, while Cornille's results tend to overestimate all lines by a few electronvolt with respect to the theoretical trend. The largest deviations in the other works are seen in the $n$ and $m$ lines, where FAC and Yerokhin values are 0.9~eV apart.

\begin{figure*}[th!]
 \centering
\includegraphics[clip=true,width=1.0\textwidth]{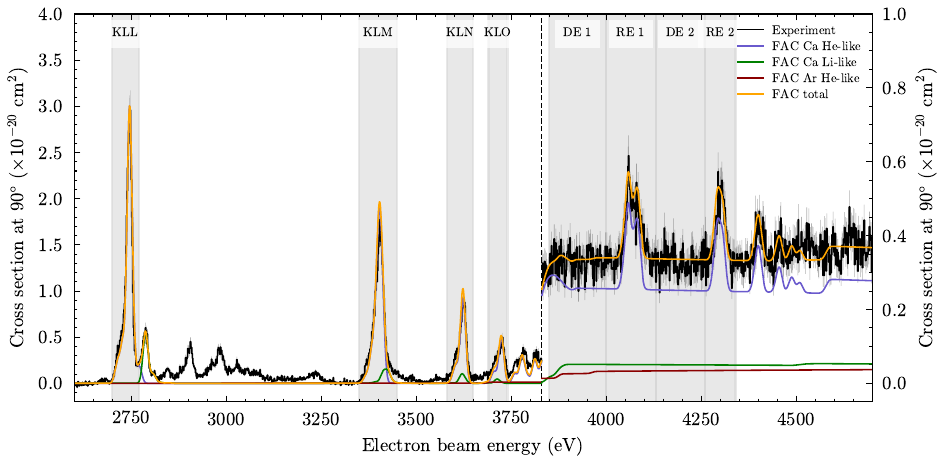}
\caption{Calibrated cross section (for photon emission at 90\textdegree) of He-like Ca DR KL$n$, DE and RE as a function of the electron-beam energy from a fast scan. Orange and green lines represent FAC values for He-like and Li-like Ca. Gray regions mark the energy ranges used to calculate the DR and RE resonant strengths and DE integrated strengths.}
\label{figs:3}
\end{figure*}
\begin{table*}[th!]
\centering
\caption{Experimental values of various Ca DR KL$n$ resonance strengths and DE and RE integrated strengths in units of $10^{-20} ~\mbox{cm}^2 \mbox{eV}$. Energy ranges used for integration for both experiment and FAC calculations are indicated in Fig.~\ref{figs:3}. The resonance strength of the KLL resonances was used for normalization. Results are compared with our FAC calculations for He-like Ca with  residual contribution of Li-like Ca and He-like Ar. Calculations from MZ and AUTOLSJ codes for He-like Ca DR structure, presented in \cite{Kato1997}, are also listed.}
\begin{tabular*}{\linewidth}{@{\extracolsep{\fill}}lcccccccc} 
\toprule
Structure & Energy region (eV) & Exp.$^\text{a}$ & FAC Total & FAC Li-like & FAC Ar He-like & FAC He-like & AUTOLSJ & MZ \\ 
\midrule
KLL	&	2700-2770	&	63 $^b$		&	62.986	&	0.107	&		&	62.879	&	69.569	&	75.225	\\
KLM	&	3350-3450	&	52	$\pm$	5	&	52.254	&	4.604	&		&	47.65	&	42.579	&	62.802	\\
KLN	&	3580-3650	&	23	$\pm$	2	&	21.291	&	2.155	&		&	19.136	&	19.949	&	24.158	\\
KLO	&	3690-3740	&	13	$\pm$	2	&	11.576	&	0.892	&	0.638	&	10.046	&	13.537	&	12.547	\\
\midrule
DE 1	&	3850-4000	&	49	$\pm$	5	&	50.038	&	6.891	&	3.837	&	39.310	&		&		\\
DE 2	&	4130-4260	&	42	$\pm$	4	&	43.316	&	6.451	&	4.417	&	32.448	&		&		\\		
RE 1	&	4000-4130	&	51	$\pm$	5	&	52.551	&	6.548	&	4.289	&	41.714	&		&		\\
RE 2	&	4260-4340	&	31	$\pm$	3	&	32.127	&	3.882	&	2.745	&	25.500	&		&		\\				
\bottomrule
\end{tabular*}
\begin{flushleft}
\footnotesize
\tablenotemark{}{Includes contributions from both He-like and Li-like charge states, with the exception of the KLL region. }\\
\tablenotemark{}{Used for cross-section calibration. }
\end{flushleft}
\vspace{1ex}
\label{tab:CaRS}
\end{table*}

\subsection{Experimental DR cross-sections}
\label{sec:cross}

Although the slow scan provided better electron beam energy resolution, we noticed changes in the charge-state distribution when comparing upward and downward scans. In contrast, these are nearly identical for the 30~ms scans, confirming that the charge-state distribution remains constant over the scan range. 
As in the slow scans, RR and DR from Ar also contaminate the fast scan, as seen in Figure \ref{figs:3} between 2800~eV and 3300~eV. A small contamination of Li-like Ca is also observable by its respective KLL resonance below 2800~eV, as well as a possible trace of Be-like Ca at 2850~eV. The Ca RR band, expected at the Ca KLL structure, is not observed due to its low cross-section, which falls below the scale of Fig.~\ref{figs:fitCa} (c). A similar situation occurs for the nearby element Ar, where the apparent Ar RR band is, in fact, dominated by N-shell RR into Ba ions (see Appendix~\ref{sec:RRveri}).

To obtain experimental cross sections, a region of interest in the two-dimensional plot containing the Ca K$\alpha$ was projected onto the electron-beam energy axis. The projected intensity was normalized to the DR KLL of our FAC calculation including the experimental energy spread and correction factors for polarization and anisotropy mentioned above. We take into account the transmission of the 1 $\mu$m-thick carbon filter in front of the SDD, as in earlier work \citep{Shah_2019, Grilo2024}. Due to the relation between emissivity and cross-sections (\cite{Grilo2024}), the summed counts were divided by $\sqrt{E_e}$, where $E_e$ is the electron beam energy. Figure \ref{figs:3} shows one of the spectra used for the determination of cross sections. We estimate the uncertainty of the KLL cross section as follow: First, the uncertainty of the theoretical calibration was obtained by considering the convergence of the CI calculation in FAC when extending the configuration space, as well as contributions from the fit to the KLL peak, in a similar manner as in \cite{Grilo2024}, resulting in calibration uncertainty in the order of $5\%$. The integrals of the total DR, DE and RE from FAC calculations at the different regions were also used to calculate different calibration factors, and the respective standard deviation contributed to the total uncertainty of around $4\%$. The statistical uncertainty of the observations contributed to the majority of the total uncertainty, accounting for a contribution of an average value of around $\sim8\%$. 
Combining all contributions, an average final uncertainty of $\sim10\%$ was obtained for the measured DR, RE, and DE cross sections.

Overall, the predicted DR features agree very well with our experimental results. Agreement is also found for the DE cross-section, and RE KM$n$ structure upon inclusion of residual contamination of Ar He-like K$\beta$ and Ca Li-like K$\alpha$ (see right side of Fig.~\ref{figs:3}). We resolve DR KL$n$ resonances up to $n=6$; beyond that, they blend together, and their resonance energies converge into the DE thresholds of the main line.

The experimental resonance strengths of each of the DR features are presented in Table \ref{tab:CaRS}. Only the KLL-, KLM-, KLN- and KLO-resonances were sufficiently isolated for a meaningful comparison with FAC predictions and other works. General agreement between the measurement and the FAC predictions for the DR structures (aside from KLL structure, which was used for normalization) is also shown with the inclusion of Li-like contamination. Although it is not possible to provide pure experimental He-like DR resonant strengths, the present FAC calculations can be compared with AUTOLSJ and MZ calculations \citep{Kato1997}. Overall, the AUTOLSJ calculations align more closely with the FAC results, showing relative deviations of around 10\%, whereas the MZ calculations exhibit deviations of approximately 20\%. 

%

\subsection{DR rate coefficients}
To directly compare our experimental results with the OPEN-ADAS database, one of the most widely used in astrophysics, both the theoretical and the experimental data were converted into rate coefficients following \citep{Gu2003, Grilo_2021}:
\begin{equation}
\alpha_{i f}^{D R}=\frac{m_e}{\sqrt{\pi} \hbar^3}\left(\frac{4 E_y}{K_B T_e}\right)^{3 / 2} a_0^3 \sum_d E_{i d} S_{i d f}^{D R} e^{-\frac{E_{i d}}{K_B T_e}}
\end{equation}
where $E_y$ is the Rydberg constant in energy units, $a_0$ the Bohr radius, $K_{B}$ the Boltzmann constant, and $T_e$ the electron temperature.
\begin{figure}[th!]
 \centering
\includegraphics[clip=true,width=1.0\columnwidth]{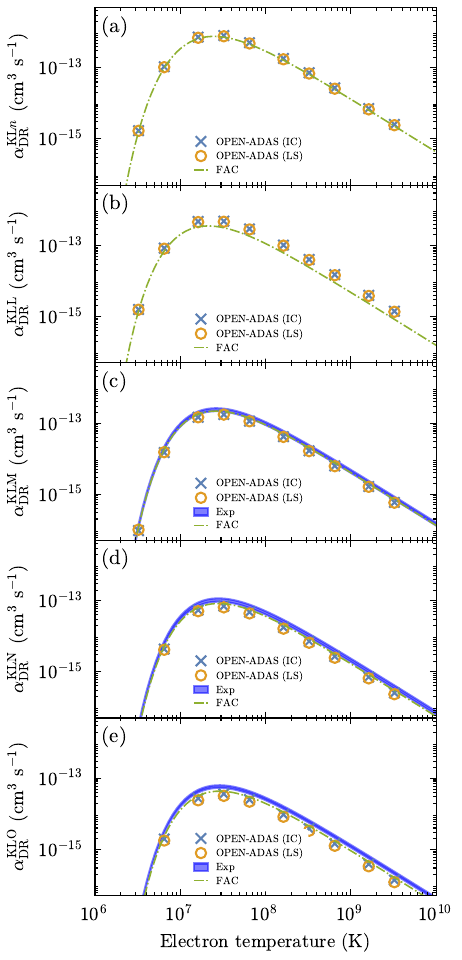}
\caption{K-shell DR rate coefficients of He-like Ca as a function of the electron temperature. Coefficients for  KL$n$ (a), KLL (b), KLM (c), KLN (d), and KLO (e). FAC data compared with OPEN-ADAS data in both IC and LS coupling schemes. Coefficients derived from the experimental data for each specific recombination structure are also presented in (c), (d) and (e).}
\label{figs:4}
\end{figure}

Figure~\ref{figs:4} compares the rate coefficients for He-like Ca from the OPEN-ADAS database with FAC values and experimental data. Due to blends with Li-like DR resonances, we add the Li-like DR resonant strengths (see Table \ref{tab:CaRS}) to the experimental uncertainty. Both IC and LS coupling schemes are represented. FAC and experimental data were grouped and compared with IC and LS data according to their main type of transitions, i.~e. the rates obtained from KLL, KLM, KLN and KLO were compared with the database data relative to $1s 2\ell 2\ell’ \rightarrow 1s^2 2\ell'$, $1s 2\ell 3\ell’ \rightarrow 1s^2 3\ell'$, $1s 2\ell 4\ell’ \rightarrow 1s^2 4\ell'$, and $1s 2l 5\ell’ \rightarrow 1s^2 5\ell'$, respectively. The experimental values of KLL are not presented, since we use them for normalization of the cross sections.

The rates from FAC are generally in good agreement with the OPEN-ADAS values in both LS and IC coupling schemes. Experimental rate coefficients, available for KLM, KLN, and KLO are also consistent with both FAC and OPEN-ADAS within experimental uncertainties. Both FAC and OPEN-ADAS slightly underestimate the experimental data, as expected from the small contamination by Li-like ions that increases the experimental resonant strengths overall. When simulating the abundance of Li-like ions for a representative temperature of $3.2\times 10^{8}~\mbox{K}$, the FAC coefficient rates deviations from the experimental data decrease from 21\% to 3\% for KLN, and from 26\% to 10\% for KLO. Considering this contribution, KLN keeps a relative error close to 10\%. The opposite effect is observed for KLL, with the databases lying higher than the FAC calculations. This is consistent with literature predictions of resonance strengths shown in Table \ref{tab:CaRS}, which are also higher than FAC values, in particular for KLL. Here, the experimental values were not included, and high-$n$ DR resonances blend with the DE threshold. When accounting for all the DR KL$n$ resonances, FAC and OPEN-ADAS agree well; at the above-mentioned temperature, the relative differences between the OPEN-ADAS IC and LS values and the ones from FAC are around 12\% and 17\%, respectively.

%

\subsection{XRISM observation}
\begin{figure}[th!]
 \centering
\includegraphics[clip=true,width=1.0\columnwidth]{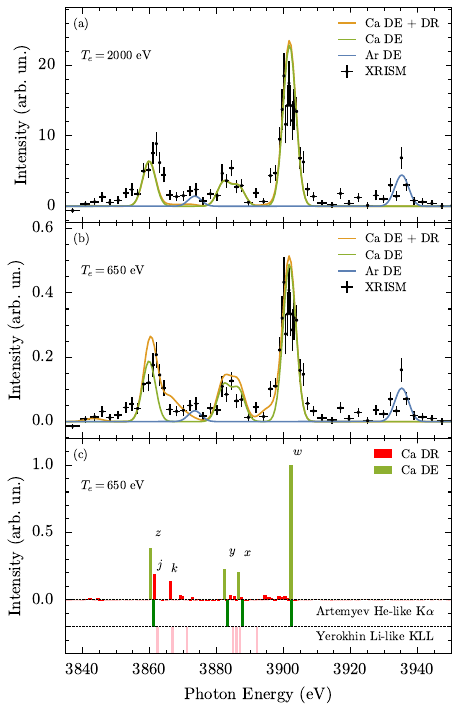}
\caption{XRISM observation of the Centaurus galaxy cluster \citep{Audard2025} (black dots) at \ion{Ca}{19} K$\alpha$ corrected for redshift in comparison with FAC-CRM simulations (solid lines) for electron temperatures T$_e$ of (a) 2000~eV and (b) 650~eV, showing only DE (green) as well as DE plus DR contributions (orange), and expected emission from H-like and He-like Ar (blue line). (c) Line emissivities at T$_e$=\,650~eV simulation. Below: He-like lines (green) and the Li-like KLL satellites (red) compared with predictions by \cite{Artemyev2005} (dark green) and \cite{Yerokhin2018} (pink), respectively.}
\label{figs:5}
\end{figure}

Several observations with XRISM show the K$\alpha$ complex of He-like Ca, as, e.~g., in the Centaurus galaxy cluster \citep{Audard2025}, where He-like Fe lines were used as a reference to infer the bulk motion of the gas at its core. Here we focus on the He-like Ca K$\alpha$ emissions around 3.9~keV. In Figure \ref{figs:5}, we compare the spectrum of a single observation with the emission spectrum simulated with a steady-state collisional-radiative-model (CRM) of the FAC package in combination with the present FAC calculations. For simplicity, only He-like and Li-like Ca ions, as well as H-like and He-like ions of Ar, were included in the model. In the first two subplots, we show FAC-CRM results for two electron temperatures (Maxwellian distribution): 2000~eV and 650~eV. Both simulations used an electron beam density of $10^{-2}$~cm$^{-3}$. These ranges of temperatures and electron density correspond to values reported by \cite{Audard2025}. We set an initial population of Ca with only He-like ions to evaluate the effects of DR satellites at different temperatures. As shown by \cite{Audard2025}, the bulk motion of the gas causes a noticeable heliocentric redshift on the observed lines. For this comparison, we shift the observational data by $E_{\small \mbox{rest}} = E_{\small \mbox{obs}}{(1 + z_r)}$, with redshift $z_r=0.0092$ to match the observed He-like Ca $w$ line to FAC. This redshift is consistent with the value $z_r	\lesssim 0.0095$ reported in the original paper.

At 2000~eV electron temperature, the overall spectrum is dominated by the $z$, $x$, $y$ and $w$ DE lines, regardless of the inclusion of DR. At a lower temperature of 650~eV, and further below the DE threshold, the Li-like satellite lines become stronger. In this regime, numerous DR features emerge as satellites of $z$, $x$ and $y$. Most of them match fairly well our simulations, and only the peak appearing near $z$ shows a noticeable deviation. 
We investigate this discrepancy in subplot (c), which shows the DE and DR emissivities from FAC-CRM, and corresponding transition energies predicted by \cite{Artemyev2005} and \cite{Yerokhin2018}. For the strongest DE and DR lines --particularly the $j$ and $k$ satellites-- FAC calculations agree well with published theoretical data, which rules out our predictions being the source of the discrepancy. Notably, the peak near the $z$ line shifts toward the $j$ and $k$ satellites, indicating that these satellites have to be carefully modeled for the analysis of observational spectra.

%
\section{Summary and Conclusions}
\label{sec:sc}
The electron-ion collisional excitation of K$\alpha$ by dielectronic recombination, direct and resonant excitation channels were studied in He-like Ca by measuring with an EBIT X-ray spectra as a function of the electron energy. We resolved dielectronic resonances up to KLR ($n=8$), and by fitting obtained their excitation energies and the photon energies emitted by these features. We compared them with FAC predictions, and found, in general, good agreement outside blends at higher-$n$ resonances. We also measured Li-like satellites of the KLL resonances and found that advanced predictions from the literature also reproduce them fairly well. In addition, we mitigate changes in the charge-state distribution using fast energy scans to also determine DR, DE and RE cross sections. The overall energy dependence, the shapes of the energy thresholds and the relative strengths of the resonances also agree well with our FAC calculations. We determined experimental resonance strengths for KLL, KLM, KLN, and KLO, and extracted DR coefficient rates based on measured and calculated cross-sections, confirming the data available in the OPEN-ADAS database for this astrophysical prominent ion.

By comparing synthetic spectra calculated from FAC, benchmarked against our measurements, with high-precision XRISM observations~\cite{Audard2025}, we highlight the role of DR satellites blending with the \ion{Ca}{19} K$\alpha$ lines. Close inspection of the region around $z$ shows that the strong $j$ and $k$ DR satellites produce a noticeable change in the amplitude and centroid of the observable $z$ line. 

The present laboratory measurements provide valuable high-resolution benchmarks of the atomic databases included in spectral fitting models incorporated in AtomDB \citep{Foster2012}, SPEX \citep{Kaastra1996} and CHIANTI \citep{Young2016}. This is essential for analyzing spectra from present and future X-ray observatories such as the XRISM~\cite{Audard2025} and ATHENA~\citep{pajot2018athena} missions, and confirming the reliability of astrophysical plasma diagnostics derived from Ca lines.

\section*{acknowledgements}
Research was funded by the Max-Planck-Gesellschaft (MPG), Germany. F.G. and P.A. acknowledge support from Funda\c{c}\~{a}o para a Ci\^{e}ncia e Tecnologia, Portugal, under grant No.~UID/FIS/04559/2020 (LIBPhys), LA/P/0117/2020 (REAL),  FCT/Mobility/1301286862/2024-25 and under contract UI/BD/151000/2021 and High-Performance Computing Chair—a R\&D infrastructure (based at the University of \'{E}vora; PI: M Avillez). C.S acknowledges support from MPG and NASA-JHU Cooperative Agreement.  


\bibliographystyle{apsrev4-2}
\bibliography{export}

\appendix

\section{Radiative recombination verification}
\label{sec:RRveri}

\begin{figure*}[th!]
 \centering
\includegraphics[clip=true,width=1.0\textwidth]{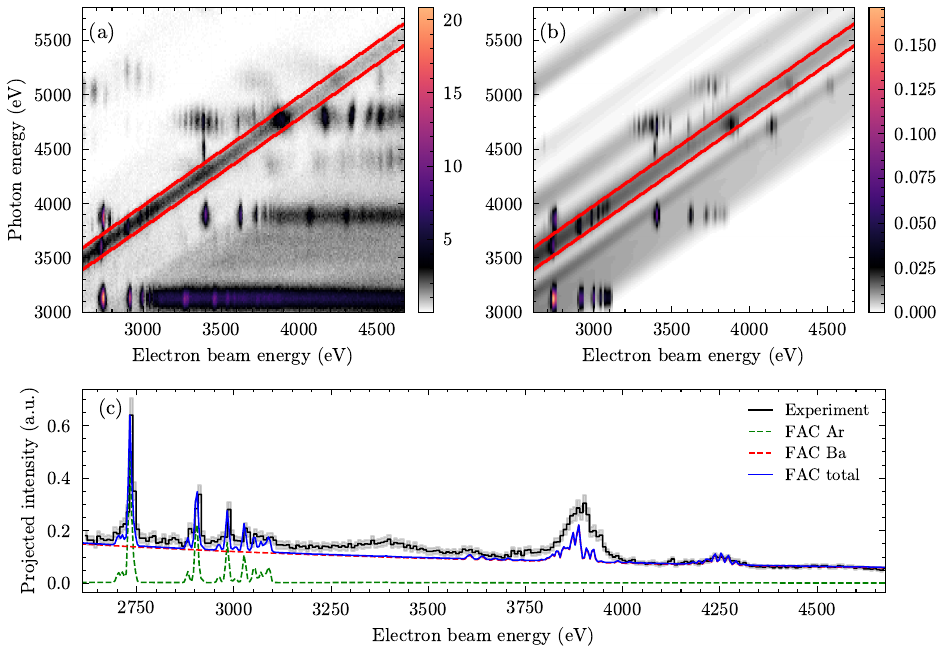}
\caption{(a): Photon yield as a function of electron-beam energy and photon energy, highlighting the RR band with a ROI. (b) Flexible Atomic Code cross sections for DR and RR of Ba, Ar and Ca. (c) Corresponding ROI counts calibrated to cross-sections.}
\label{figs:RR}
\end{figure*}

As already mentioned in Sec.~\ref{sec:cross} and illustrated in Fig.~\ref{figs:fitCa} (a) and (c), the Ca RR band into the L-shell is too weak to be detected in the short exposure time spent at each energy during the scans, and the feature appearing at the location of the Ar RR band is actually dominated by N-shell RR of Ba ions. 
Barium, originating from the coating of the EBIT cathode, is commonly present in EBIT observations. To minimize its influence, the trap is periodically dumped, purging all ions. This is effective because Ba ions fill up more slowly than the purposely injected Ca, and it takes approximately 13~s for the Ni-like Ba$^{28+}$ (ionization energy of 970~eV) charge state to appear. This estimate is based on population-balance equations involving only RR and collisional ionization processes from a previous study \citep{Grilo_2021}. We set the dump cycle to 33~s, as similar simulations with the same code indicate that this is required to produce 90\% of He-like Ca. Therefore, the presence of [Ar]$3d^{m}$ ($m=0,...10$) charge states of Ba is expected at the EBIT trap. 
Figure~\ref{figs:RR} shows an extended photon emission map of Fig.~\ref{figs:fitCa} (a) together with FAC predictions (b), as well as the counts (c) for the RR ROI indicated there. 
 Here, the experimental counts have been calibrated to match the area under the Ar KLM peak, based on FAC calculations of the same atomic process, including L-shell RR of He-like Ar and N-shell RR and LM$n$ resonances of Ba ions. However, a comprehensive analysis of the complex [Ar]$3d^{10}$ DR LM$n$ structure of Ba observed at higher energies is beyond the scope of this work. Nonetheless, these observations show how the blended-in RR band originates from these ions. Accordingly, our FAC calculations are limited to configurations with up to two unpaired electrons and holes in the $3d$ shell. The relative populations of the selected Ba ions used in our predictions are estimated based on population-balance simulations for the dump cycle. This approach effectively accounts for the missing charge states, as all relevant ions exhibit similar RR cross sections.
 Due to the much higher charged states reached by the Ba ions under the present conditions, their RR cross-sections are much larger than those of the Ar HCI, and therefore Ba dominates the RR band region under the present experimental conditions. 

\section{Photon energy calibration}
\label{sec:calibration}
\begin{figure*}[th!]
 \centering
\includegraphics[clip=true,width=1.0\textwidth]{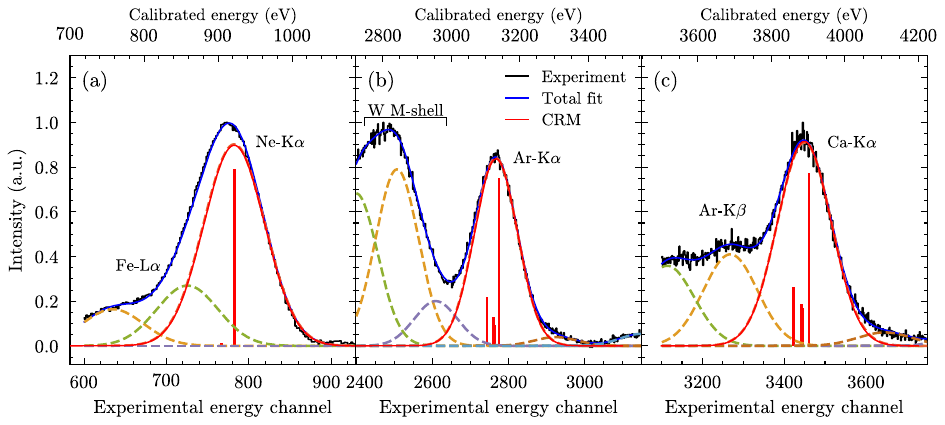}
\caption{Multiple Gaussian fits performed to the K$\alpha$ line complexes of He-like Ne (a), Ar (b) and Ca (c). Experimental data (blue curve) and total fit (red dashed lines) shown versus uncalibrated detector energy. Top axis: calibrated photon energy. The relative emissivities of the K$\alpha$ lines are represented by the vertical red bars, as well as the corresponding Gaussian convoluted emission profile (solid red line).}
\label{figs:fits}
\end{figure*}

To convert detector channels into physical photon energies, a linear calibration was performed using a selection of soft X-ray lines spanning a wide spectral range. Only well-known reference energies from He-like K$\alpha$ transitions from Ne, Ar and Ca having clearly resolved profiles were included in the fit.
Argon lines present due to a small leak were also used for the calibration. After the main observations, Ne was injected and excited under conditions similar to those of the main experiment. We also detect background lines from H-like and He-like O, Ne-like Fe, and several charge states of Ba and W (from EBIT cathode), but do not use them for calibration. Line centroids from Gaussian fits with associated uncertainties were obtained. The fit of the three K$\alpha$ line complexes used for the calibration is displayed in Figure~\ref{figs:fits}. We use a shared width for all the Gaussians in each fit and subtract a constant background. In the case of Ar, the corresponding line is well isolated, yielding a very low statistical error. Below the main Ar feature, background lines attributed to Co-like to Cu-like M-shell emission of W ($n_i=4,5,6 \rightarrow n_f=3$) \citep{Clementson2010} appear. For Ne, we notice some asymmetries in the low energy tail of the main Gaussian. In this case, a shift of the centroid observed in the overall best fit seems to be due to a second Gaussian peak nearby. Such Gaussians arise from known contamination by H-like O and Ne-like Fe, leading to K$\beta$ and L$\alpha$ emission around $\sim 775$~eV and $\sim 850$~eV, respectively. A higher uncertainty for the centroid of the Ne K$\alpha$ feature was taken into account. On the left side of the main Ca K$\alpha$ lines are features attributed to the K$\beta$ lines of Ar, around $\sim 3680$~eV.

Table~\ref{tab:calib} lists measured detector positions (in channels) and corresponding reference energies used for calibration. Only transitions with reliable literature values compiled at the NIST ASD (except for the Ca lines) were used for calibration. The  Ne lines at ADS NIST are theoretical values retrieved from \cite{Johnson2002}, and for Ar and Ca from \cite{Artemyev2005}.
For the final calibration line of each element, an effective value for the K$\alpha$ line complex was calculated by fitting the emission spectrum obtained with a steady-state FAC CRM for the experimental conditions, with the respective transition energies replaced by those used for calibration and emissivities given by the FAC calculations. The weights obtained for each line are reported in Table ~\ref{tab:calib}. Energies for $z$, $x$, $y$ and $w$ were also calculated with FAC, and agree within 1.4 eV with the theoretical ones used for the calibration, with the FAC values generally underestimating the values from \cite{Artemyev2005} and \cite{Johnson2002}. For the cases of Ar and Ca, the FAC $w$ values agree with the QED-based values of \cite{Artemyev2005} within $<$0.15~eV. Several laboratory observations of the K$\alpha$ lines for these elements were also compiled in the table for reference. Although high-precision laboratory observations are known for Ar and Ca, to the best of our knowledge, only wavelength measurements of Ne lines from astrophysical observations are currently available. Only the $w$ line has been reported as the total K$\alpha$, which might be attributed both to the energy closeness of the lines and to the expectation that under those plasma conditions, other He-like lines are suppressed. 

\begin{table*}[th!]
\centering
\caption{Experimental photon energies (channels), corresponding He-like energies (eV) compiled at the NIST database, and relative intensities used in the calibration.  The values of K$\alpha$ presented in the table reflect the centroid value of fits to synthetic spectra generated for the calibration lines with their relative intensities. The NIST calibration lines are taken from accurate theoretical calculations from \cite{Johnson2002} for Ne, from \cite{Artemyev2005} for Ar and a.  Their relative intensities were calculated with the FAC-CRM package for the EBIT conditions for each individual element. Measurements of each He-like line are also listed with the respective uncertainty given in parentheses. \label{tab:calib}}
\begin{tabular*}{\linewidth}{@{\extracolsep{\fill}} l l
                S[table-format=4.2]
                c
                S[table-format=4.2]
                S[table-format=4.2]
                l S[table-format=4.2]}
\toprule
Element & Transition
& {Experimental}
& {Calibration lines} & {FAC} 
& {Line measurements} & Ref. & {Rel. Intensity} \\
 & & {(arb. un.)} & {(eV)} & {(eV)} & {(eV)} & & \\
\midrule
Ne  & $w$ & {} & 922.09 & 921.99 & 922.124(34) & \citep{Liao2013}$^\text{a}$& 0.979 \\
&& {} & &  & 922.148(82) & \citep{Yao2009}$^\text{b}$ &  \\
    & $x$ & {} & {n.l.}$^\text{c}$  & 914.38 & & & <0.001 \\
    & $y$ & {} & 914.68 & 914.19 & & & 0.005 \\
    & $z$ & {} & 905.06 & 904.03 & & & 0.016 \\
    & K$\alpha$ & 782.30 & 921.78 & &  & \\
    \midrule
Ar  & $w$ & {} & 3139.58 & 3139.49 & 3139.5927(76) & \citep{Machado2018} & 0.65 \\
    & & & & & 3139.5810(92) & \citep{Kubicek2014} & \\
    & & & & & 3139.552(37) & \citep{Deslattes1984} & \\
    & & & & & 3139.57(25) & \citep{Briand1983} & \\
    & $x$ & {} & 3126.28 & 3125.54 & 3126.283(36) & \citep{Deslattes1984} & 0.08 \\
    &     & {} &         & & 3126.37(40) & \citep{Briand1983} &  \\
    &     & {} &         & & 3128(2) & \citep{Dohmann1978} &  \\
    & $y$ & {} & 3123.53 & 3122.82 & 3123.521(36) & \citep{Deslattes1984}  & 0.11 \\
    &     & {} &         & & 3123.57(24) & \citep{Briand1983}  &  \\
    & $z$ & {} & 3104.15 & 3103.01 & 3104.1605(77) & \citep{Amaro2012} & 0.19 \\
    & K$\alpha$ & 2771.22 & 3130.26 & & & & \\
    \midrule
Ca  & $w$ & {} & 3902.38 & 3902.25 & 3902.19(12) & \citep{Rice2014} & 0.55 \\
    &     & {} &         & & 3902.43(18) & \citep{Aglitsky1988} &  \\
    & $x$ & {} & 3887.76 & 3886.98 & 3887.63(12) & \citep{Rice2014} & 0.13 \\
    & $y$ & {} & 3883.32 & 3882.58 & 3883.24(12) & \citep{Rice2014} & 0.12 \\
    & $z$ & {} & 3861.21 & 3860.04 & 3861.11(12) & \citep{Rice2014} & 0.19 \\
    & K$\alpha$ & 3449.55 & 3890.20 & & & & \\
\bottomrule
\end{tabular*}
{\begin{flushleft}
\footnotesize
\tablenotemark{}{From 36 Chandra High-Energy-Transmission-Grating observations. Assuming no line contamination from other He-like lines. The CODATA 2018 \citep{codata2018} recommended value of $hc=1.23984198\times 10^{-6}$\,eV/m was used for conversion.}\\
\tablenotemark{}{Same as previous, but for the interstellar medium surrounding Cyg X-2.}\\
\tablenotemark{}{Not listed in NIST ADS and \citep{Johnson2002}.}\\
\end{flushleft}
}
\vspace{1ex}
\end{table*}

\begin{figure*}[th!]
 \centering
\includegraphics[clip=true,width=1.0\textwidth]{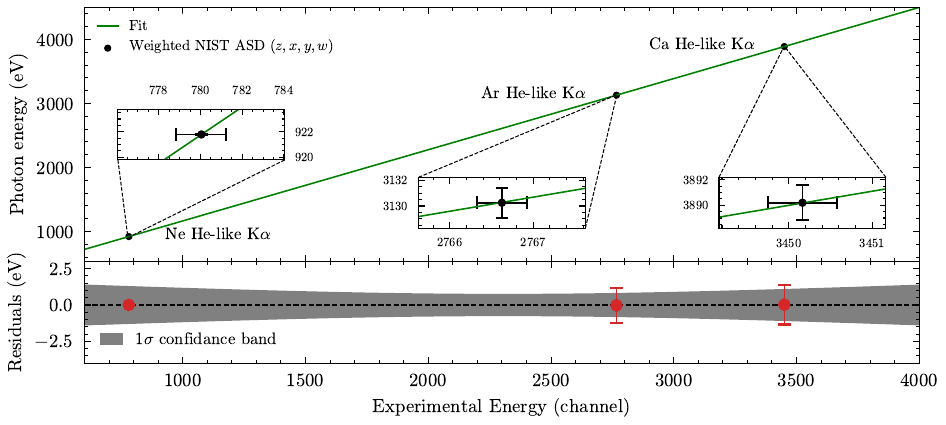}
\caption{Linear calibration fit using lines from Ne, Ar and Ca. Error bars: uncertainties in measured line centroids and theoretical energies for calibration (see insets); fit residuals shown at the bottom.}
\label{figs:SDD}
\end{figure*}

A least-squares regression weighted by uncertainties yielded the linear calibration function shown in Figure~\ref{figs:SDD}. The abscissa uncertainties are taken from those of the fitted centroids, which presented a fit uncertainty of around 1.2~eV for Ne and less than 0.5~eV for Ar and Ca, while for the ordinate, the uncertainties from using FAC-CRM to weight the K$\alpha$ lines dominate. To estimate them, a conservative value of 20\% was then used to vary the FAC-CRM line emissivities for each line. The highest changes of the effective centroid were taken as uncertainties for the transition energies of the calibration lines. In Ar and Ca the K$\alpha$ complex partially blends the $x$, $y$ and $z$ lines, and the total uncertainty goes slightly over 1~eV. In Ne, the K$\alpha$ complex is dominated by $w$, and the centroid variations introduce an error of  $<$0.1~eV. The calibration line presented in Figure~\ref{figs:SDD} has residuals of 0.1~eV for Ar and Ca, and 0.01~eV for Ne. This makes us confident in the linear calibration of the SDD detector. We also noted that using the measured values of He-like lines listed in Table~\ref{tab:CaRS} for calibration instead of NIST values did not change either the linear calibration function or its uncertainty. The final uncertainty in the photon calibration includes the statistical uncertainty from the fit of the experimental reference lines, as well as the estimated uncertainty from the theoretical reference lines.
The $1\sigma$ confidence interval for the calibration fit is represented as the uncertainty band in the residuals shown in Figure~\ref{figs:SDD}. This, combined with the centroid uncertainty from the peak fitting, yields the total uncertainties in experimental photon energies shown in Tables \ref{tab:CaLiComp} and \ref{tab:Cafit}.

\end{document}